# Non-Euclidean conformal devices with continuously varying refractive index profiles based on bi-spheres

Wenjing Lv[1], Jiaojiao Zhou[2], Y. Liu (刘泱杰)[3, 4*] and Lin Xu[1†]

[1]*Information Materials and Intelligent Sensing Laboratory of Anhui Province & Institutes of Physical Science and Information Technology, and Center of Free Electron Laser & High Magnetic Field, Anhui University, Hefei 230601, Anhui Province*

[2]*Department of Mathematics and Physics, and Key Laboratory of Advanced Electronic Materials and Devices, Anhui Jianzhu University, Hefei 230601, Anhui Province*

[3]*Department of Physics, School of Physics, Hubei University, Wuhan 430062, Hubei Province*

[4]*Department of Optics & Optical Engineering, School of Physical Sciences, University of Science and Technology of China, Hefei 230026, Anhui Province*

Corresponding authors: * yangjie@hubu.edu.cn, † xuin@ahu.edu.cn

**Abstract**: Either conformal transformation optics or geodesic mapping provides a design method to bend light rays in two-dimensional space with a nonuniform refractive index profile. In this paper, we combine both methods above to design a conformal invisible cloak based on bi-spheres with a refractive index profile varying from 0 to 10.7, smaller than 24.6 for the previous case of a single sphere. Moreover, we obtain an omnidirectional retro-reflector and a specular reflector by making position adjustments to mirrors, and achieve similar invisible effect by tuning sizes of the bi-spheres. Our work expands the toolkits for designing conformal devices with continuously-varying index profile.

## 1. Introduction

From the geometric perspective of gravity in general relativity, the mass-energy distribution is viewed as the source of the curvature of space-time [1]. This geometric notion therefore inspired the transformation optics [2, 3] to treat the electromagnetic materials as the curvature of electromagnetic space, which gives an intuitive recipe for designing versatile optical devices. Due to the deformations induced by coordinate transformations, the electromagnetic permittivity and permeability tensors in transformation-designed devices exhibit inherent inhomogeneity and anisotropy, which hinders their pragmatic realization. With the development of metamaterials [4-7] made of artificially structured units, a two-dimensional (2D) reduced version of the invisible cloak was fabricated with split-ring resonators [8]. Thereafter, various transformation-designed and metamaterials-fabricated devices were reported, including carpet cloaks [9, 10], field rotators [11, 12], field concentrators [13, 14], field shifters [15, 16], transmuted singular devices [17, 18], and optical illusion devices [19, 20]. The comprehensive details are available in the reviews [21-24]. Such an electromagnetic paradigm to design new functional devices also applies well to other wave realms [25-27]. More exhaustive development of transformation optics can be found in its roadmap [28].

As a special branch of transformation optics, conformal transformation optics [2, 29] provides an analytic toolkit for determining nonuniform, isotropic refractive index profile, to control light rays precisely in 2D space. In the very first design of optical invisible based on conformal transformation optics [2], the guiding profile of refractive index placed in the upper Riemann sheet of the analytic mapping was discontinuous along the branch cut to the lower Riemann sheet, which results in unavoidable reflections and compromises its perfect cloaking effect. Such a discontinuous index issue was leveraged by adding a Mikaelian lens as a guiding refractive index profile to design for a transparent device with an optimized index range [30, 31]. This issue was resolved because the added profile make up for the discontinuity along the branch cut [30]. It was also noted that the index-guiding profile can be mapped to a non-Euclidean surface with a uniform index profile under the concept of the geodesic mapping [32, 33]. Furthermore, inspired by the broadband property of non-Euclidean cloaking [34], conformal transparent and invisible devices were achieved with continuously-varying refractive index profile based on the composition of the conformal mapping and the geodesic mapping from a 2D non-Euclidean space [35], significantly improving the cloaking effect.

In this paper, we further design conformal devices with non-Euclidean geometry in the framework of conformal transformation optics with geodesic mapping, which may yield practical optical devices with a continuously-varying refractive index profile. In Section 2, we construct a non-Euclidean virtual space with equisized bi-spheres and mirrors to create a conformal invisible cloak for both light rays and waves at eigenfrequencies. The required refractive index profile varies continuously from 0 to 10.7, smaller than 24.6 for the previous case of a single sphere [35]. By reorienting the mirrors from horizontal to vertical, we also achieve an omnidirectional retro-reflector in Section 3. With both horizontal and vertical mirrors present, it works as a specular reflector in Section 4. In Section 5, we demonstrate that the cloaking effect for light rays remains even when the sizes of the bi-spheres differ. Finally, we conclude in Section 6.

**2. Non-Euclidean conformal invisible cloak based on equisized bi-spheres**

Let's revisit the method of combining the conformal mapping and the geodesic mapping to map a non-Euclidean space with a uniform refractive index profile to a physical space with a nonuniform refractive index profile. Conventionally, conformal mapping [2, 29] is an analytical function [36] that can link points between $z$ complex plane (physical space) and $w$ complex plane (virtual space). The angle-preserving condition leads to the following relation between the refractive index profiles $n(z)$ and $n'(w)$ in both virtual and physical spaces,

$$n(z) = \left|\frac{dw}{dz}\right| n'(w). \qquad (1)$$

This relation indicates that two planes with refractive index profiles satisfying Eq. (1) are optically equivalent for light rays. In fact, 2D surfaces are all conformal flat with the metric tensor components $g_{ij} = n^2 \delta_{ij}$, where $\delta$ is the Kronecker delta symbol, $i$ ( $j$ ) is the tensor rank index and $n$ is the scaler function [37]. Such a scaler $n$ can be treated as the refractive index profile for light rays traveling on the surfaces. Thus we can calculate light trajectories by solving the geodesic equation on curved surfaces.

The coordinate transformation between curved surfaces is referred as geodesic mapping in this paper. This widely utilized geodesic mapping [32, 33] works from a plane lens with rotationally symmetric refractive index profile $n(r)$ to a surface of revolution with uniform refractive index profile $n'(h) = 1$, which is written as

$$\rho = n(r)r \quad \text{and} \quad dh = n(r)dr, \qquad (2)$$

where $\rho$ represents the radial coordinate, and $h$ denotes the length measured along the meridian from the north pole on the surface of revolution. We will employ this geodesic mapping in the subsequent design which maintains 2D rotational symmetry.

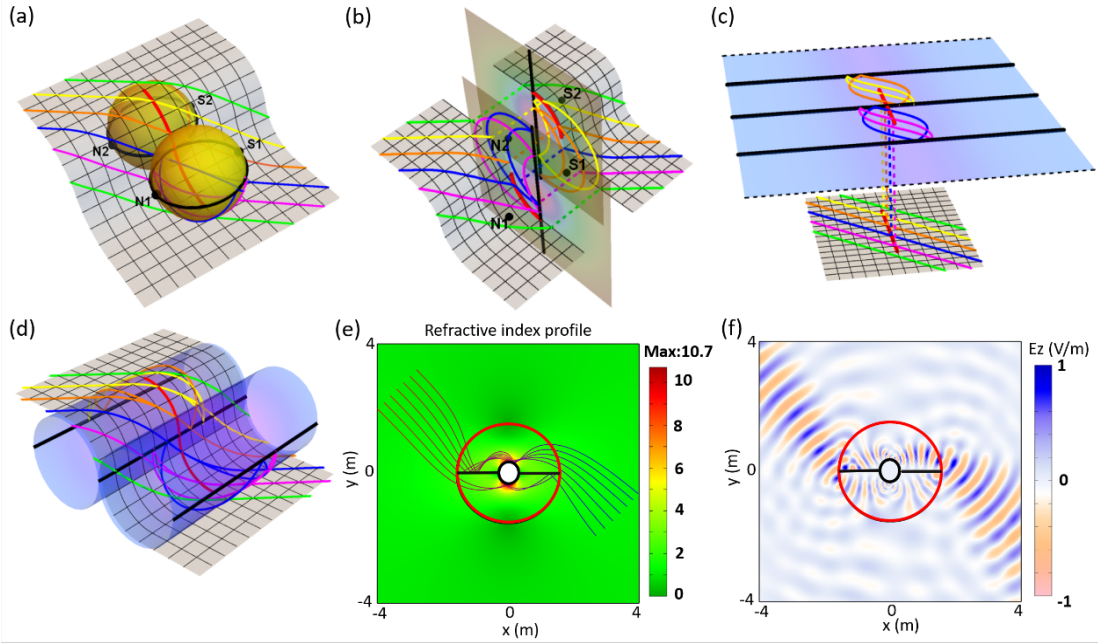

**Fig. 1. Conformal invisible cloak based on equisized bi-spheres**. (a) The Non-Euclidean virtual space comprises a meshed planar surface with equisized bi-spheres connected by a bifurcated red line and the two mirrors are indicated by solid black circles. (b) By geodesic mapping, bi-spheres are mapped to two Maxwell's fish-eye lenses on two distinct vertical planes, deliberately separated to enable visualization of the light rays on each plane. (c) By exponential conformal mapping, the two vertical planes in (b) are mapped to the two ribbon-shaped regions bounded by a dashed black and the middle solid black line. (d) An alternative intuitive representation of (b) involves curling the two ribbon-shaped regions into two cylinders. By conformal dual-logarithmic mapping with a linear term, the upper and lower sheets in virtual space in panel (d) are mapped to physical space of the invisible device with refractive index profile ranging from 0 to 10.7. The red curve and thick black lines represent images of the branch cut and mirrors, respectively. (f) A Gaussian beam impinges on the device at the angle of π/4 rad.

We now present a design of the conformal device based on equisized bi-spheres. As shown in Fig. 1(a), a non-Euclidean virtual space ($w_2$) is constructed comprising a meshed surface and equisized bi-spheres. The meshed plane is intrinsically flat and can be represented by a complex variable, to which bi-spheres are affixed along a red line that bifurcates into two red quarter arcs with the radius $r_0$. On the bi-spheres, the two black closed circles through the north and south poles are mirrors for incident light, which can be simulated by the boundary condition of perfect electric conductor [38]. Light rays with different colors in Fig. 1(a) propagate on the plane at the incident angle π/4 toward the red line. When the light rays meet on the red line, they enter the bi-spheres and travel along the geodesic lines (great circles). After being reflected by the mirrors twice, the light rays return to the red line, resuming their original positions and directions. They then continue to propagate further along straight lines on the plane. Rays that avoid the red line follow the straight paths on the plane, as indicated by two green lines. In such non-Euclidean virtual space, two mirrors prevent light rays from reaching into the two half-spheres. When observed away from the bi-spheres, all light rays appear to propagate in the original direction as if the bi-spheres and mirrors are invisible.

Next, we shall perform three mappings for the virtual space depicted in Fig. 1(a) in sequence to design our device, which respectively are geodesic mapping described in Eq. (2) in Fig. 1(b), exponential mapping in Fig. 1(c) and dual-logarithmic mapping in Fig. 1(e). First, we exclusively apply the geodesic mapping to each of bi-spheres, resulting in two Maxwell's fish-eye lenses with refractive index profiles $n(w_1) = 2/(1+(|w_1|/r_0)^2)$ on the two vertical planes, as illustrated in Fig. 1(b). The two vertical planes are deliberately spaced apart to visualize light rays on them. The south poles S1/S2 in Fig. 1(a) are projected to the centers of two vertical planes in Fig.

1(b), while the north poles N1/N2 in Fig. 1(a) to the infinities of two vertical planes in Fig. 1(b). The two solid black circles in Fig. 1(a) are mapped to two vertical solid black lines in Maxwell's fish-eye lenses in Fig. 1(b). In this paper, the non-Euclidean geometry is chosen as bi-spheres, making the associated geodesic mapping to be the so-called stereographic projection from north poles. In general, other geodesic mapping also works for non-spheres with rotational symmetry. Subsequently, we employ exponential conformal mapping $w_1 = \exp(w)$ to map two Maxwell's fish-eye lenses on vertical planes in Fig. 1(b) to two ribbon-shaped regions in Fig. 1(c). In each ribbon-shaped region located between the dashed black line and the middle solid black line (crossing the red line), a truncated Mikealian lens of a width $4\pi$ is used, with the refractive index profile $n(w) = 1/\cosh(\mathrm{Re}(w)/4)$. For visual clarity, we coil the two ribbon-shaped regions into two cylinders with the perimeter $4\pi$, as illustrated in Fig. 1(d). Using the dual-logarithmic mapping with a linear term $w = z + \alpha \log(z - \beta) - \alpha \log(z + \beta)$ [30], we map the Riemann surface in Fig. 1(c), which consists of an infinite complex plane (lower sheet) and the ribbon-shaped plane (upper sheet), to physical space represented in Fig. 1(e). Here parameters $\alpha$ and $\beta$ are chosen as 4 and 0.3125, respectively. Under this dual-logarithmic mapping, the red branch cut is mapped to the circle-like red closed curve, and the three straight mirrors are mapped to a small circle-like black closed curve attached by two straight black lines, as depicted in Fig. 1(e).

Thus from a non-Euclidean virtual space and then using three mappings described above, we design a conformal invisible device with the continuously varying refractive index profile ranging from 0 to 10.7 as illustrated by the background contour plot in Fig. 1(e). The expression of this index profile, based on Eq. (1) and (2), is

$$n(z) = \begin{cases} \left|1 + \dfrac{2\alpha\beta}{z^2 - \beta^2}\right|, & \text{outside red curve} \\ \left|1 + \dfrac{2\alpha\beta}{z^2 - \beta^2}\right| \cdot \dfrac{1}{\cosh(\mathrm{Re}(w)/\alpha/2)}, & \text{inside red curve} \end{cases}. \quad (3)$$

The numerical result in Fig. 1(e) from COMSOL Multiphysics demonstrates that parallel light rays incident on the device are bent by nonuniform refractive index profile and reflected by mirrors twice, rendering the region inside the small black closed curve undetectable. The propagation of a Gaussian beam also reveals the trajectories of light rays of our design in geometrical optics, as shown in Fig. 1(e) and 1(f).

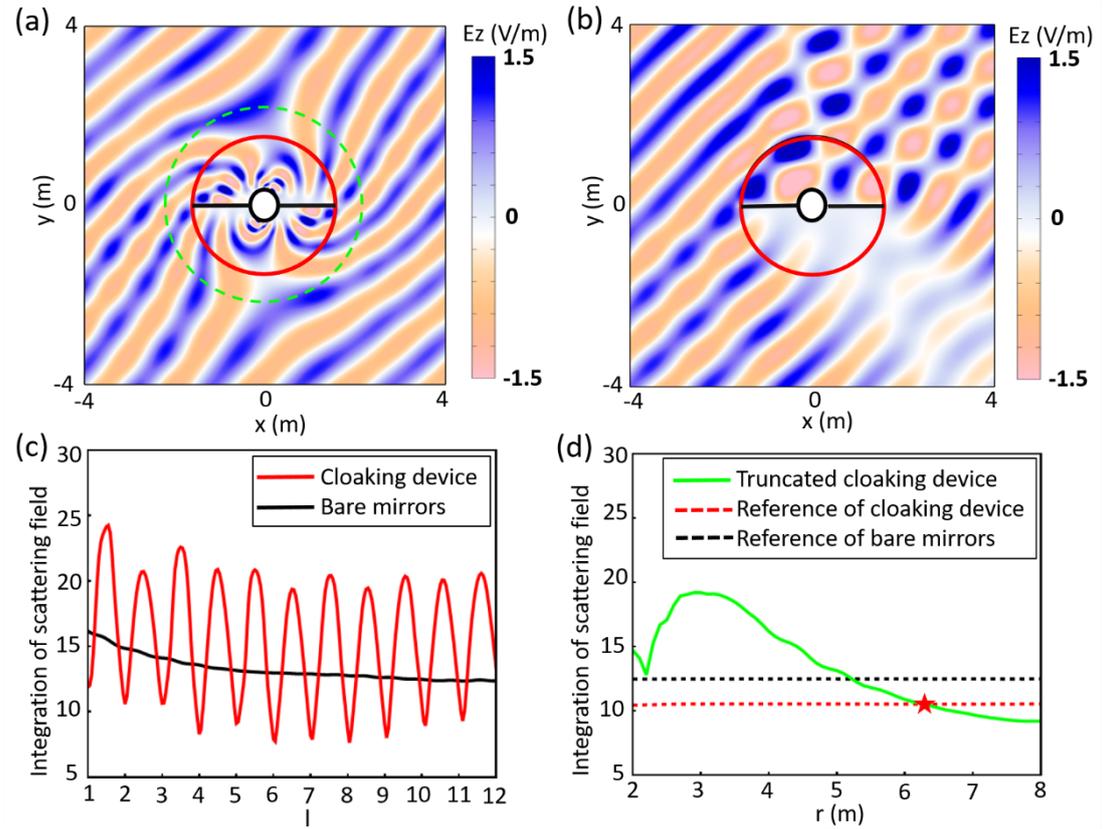

**Fig. 2. Cloaking effect at eigenfrequencies.** A plane wave with $l=10$ impinges on the cloaking device (a) and the bare mirrors (b) with an incident angle of π/4 rad. (c) The integration of scattering field along the circle with the radius 8 m for the cloaking device and the bare mirrors at different wave lengths represented by $l$ from Eq. (4). (d) The integration of scattering field along the circle of radius 8 m for the truncated cloaking device varies with the truncation radius $r$, denoted by the dashed green circle in (a), at the wavelength with $l=10$. The reference value of the integration of scattering field for the cloaking device and the bare mirrors are illustrated with red and black dashed lines, respectively.

Moreover, conformal transformation optics is applicable not only for geometric optics but also under discrete eigenfrequencies for wave optics [39]. Since waves accumulate an additional phase when propagating on bi-spheres in Fig. 1(a), the broken phase at the branch cut generally disrupts the cloaking effect. Nevertheless, at discrete frequencies, the additional phase can be a multiple of $2\pi$, resulting in a undetectable effect. The discrete frequencies for waves on the sphere are associated with spherical harmonic functions, which are written as

$$\lambda = \frac{2\pi r_0}{\sqrt{l(l+1)}}, \qquad (4)$$

where $l$ is any positive integer. For a plane wave with $l=10$ impinging on the cloaking device, the plane wavefront is preserved when it leaves the red circle as shown in Fig. 2(a). Compared with the scenario depicted in Fig. 2(b), the plane wavefront becomes distorted after encountering the bare mirrors, producing a clearly observable reflection pattern. This comparison demonstrates the effectiveness of the conformal invisible cloak with a continuously varying refractive index profile, which is designed based on two tangent spherical geodesic lenses. The integration of scattering field, defined by the integration of absolute value of scattering electric field on the circle with radius $2\alpha$, namely 8 m here, are plotted for both the cloaking device and the bare mirrors in Fig. 2(c). This phenomenon indicates that the scattering of our devices at integer multiples of $l$ are significantly reduced, consistent with our previous studies [30, 35]. These reductions of scattering fields at eigenfrequencies can perform better if we choose a larger integral circle in the numerical simulation. It is known that conformal transformation optics leads to inhomogeneous refractive index profile occupying the entire plane. Inspired by previous approaches [40, 41] to characterizing the scattering of a cloaked object, we further quantify the performance of our cloaking device under truncated circle with the radius $r$, as shown in the dashed green circle in Fig. 2(a), by the integration of scattering field along the circle with the radius 8 m at the wavelength with $l=10$ as shown in Fig. 2(d). Comparing to the reference value of the integration of scattering field for the cloaking device and the bare mirrors, it turns out that the truncated cloaking device with the radius larger than 6.3 m has good cloaking effect. This is indicated by the red asterisk intersected by the solid green line and the dashed red line.

We have successfully demonstrated the invisible effect of the conformal device based on bi-spheres at the geometric optics and wave regimes. The designed index profile can be explicitly expressed in Eq. (3). In Sec. 3, we will slightly change the position of the mirrors to achieve an omnidirectional retro-reflector and a specular reflector. And in Sec. 4, we adjust the size of the bi-spheres allows for designing various invisible devices.

### 3. Omnidirectional retro-reflector

To create an omnidirectional retro-reflector, we reconfigure the bi-spheres and their connection to the meshed planar surface in non-Euclidean virtual space, as depicted in Fig. 3(a) in comparison to Fig. 1(a). We attach the entire red branch cut to the half equator of the lower sphere, rather than attaching two quarter equators in both spheres in Fig. 1(a). While the meshed planar surface exhibits some differences here, it remains flat as depicted in Fig. 3(a). Since there is no contact point between the upper sphere and the meshed planar surface, it kisses the lower sphere at the symmetric point $K$, corresponding to points $K$ and $K'$ in the following figures. Moreover, we position the longitude $\widehat{N_1KS_1}$ and the half-equator as intersecting mirrors depicted within two black solid half-circles. In this non-Euclidean virtual space, any light rays reflected by mirrors within the lower sphere return along their incident directions without entering the upper sphere, as exemplified by the blue light ray depicted in Fig. 3(a). Consequently, the upper sphere remains inaccessible to external light rays, resulting in the invisibility of the upper sphere. We also designate the longitude $\widehat{N_2KS_2}$ in the upper sphere as the mirror, which can be combined with the longitude $\widehat{N_1KS_1}$ to correspond to the line mirrors passing through points $K$ and $K'$ as depicted in Fig. 3(b). The dashed longitude , together with the solid $\widehat{N_2KS_2}$, forms a great circle dividing the entire upper sphere into two halves, which are mapped to the right Maxwell's fisheye in Fig. 3(b).

Utilizing the same geodesic mapping from Fig. 1(a) to 1(b), bi-spheres in Fig. 3(a) are projected onto two vertical Maxwell's fisheye planes in Fig. 3(b). When the blue light ray traverses the red branch cut to the left Maxwell's fisheye plane, it follows the guidance of the inhomogeneous refractive index profile, reflects twice off the mirrors, and returns to its original impinging direction without noticing the right Maxwell's fisheye plane. If the light rays entre from right, the branch cut will appear on the right Maxwell's fisheye plane.

By employing the same exponential mapping from Fig. 1(b) to 1(c), the left Maxwell's fisheye plane in Fig. 3(b) is projected onto the middle ribbon-shaped region between two black solid lines on the upper sheet in Fig. 3(c). Meanwhile, the right Maxwell's fisheye plane, intersected by the solid and the dashed rays, is mapped to two ribbon-shaped regions, each bounded by a solid and a dashed line. The blue light ray will be guided back to the opposite propagating direction by the upper sheet of the Riemann surface in Fig. 3(c). For visual clarity, similar to Fig. 1(d), we coil the middle ribbon-shaped region into the lower cylinder in Fig. 3(d), while two ribbon-shaped regions remaining on the sides are combined to coil into the upper cylinder.

Under the same dual-logarithmic mapping from Fig. 1(c) to 1(e), we achieve an omnidirectional retroreflector in Fig. 3(e), with a refractive index profile identical to as depicted in Fig. 1(e). The only distinction is the replacement of two horizontal black mirrors depicted in Fig. 1(e) with two vertical black mirrors in Fig. 3(e). The parallel light rays will reflect back to their original direction, with only a slight transverse sidestep. The retro-reflection effect under numerical simulation for a Gaussian beam is depicted in Fig. 3(f).

It is evident that the non-Euclidean geometry components and mappings are the same for the invisible cloak in Fig. 1 and the omnidirectional retro-reflector in Fig. 3. In the subsequent sections, we continue to illustrate the theme of our paper that the mirror designs can be applied to a variety of optical devices with in the same refractive index profile, thus expanding application ranges for conformal transformation optics.

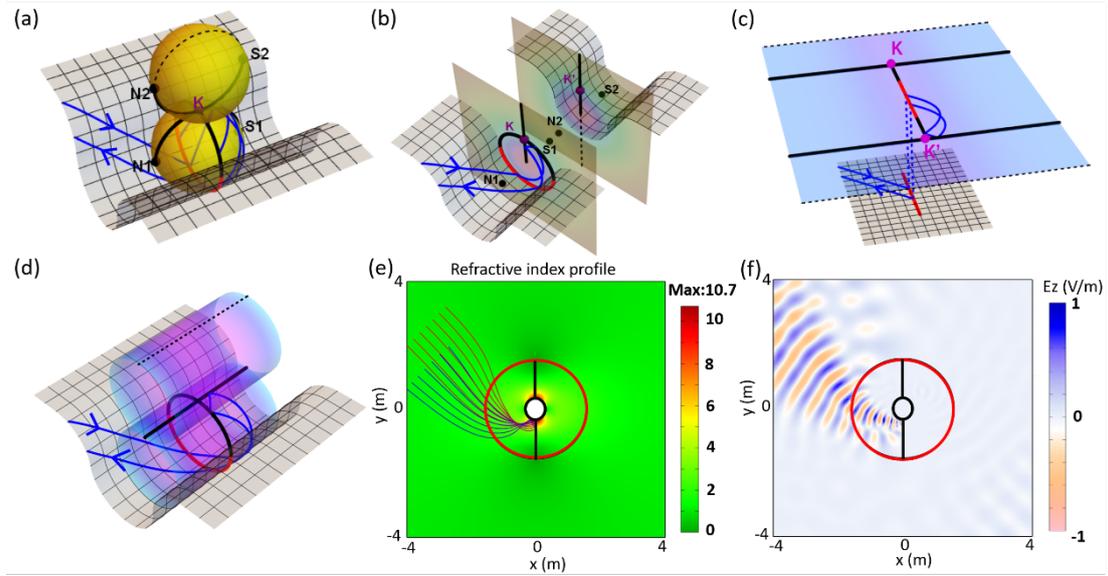

**Fig. 3. Omnidirectional retro-reflector based on bi-spheres**. (a) Non-Euclidean virtual space. (b) The intermediate space after geodesic mapping. (c) Riemann surface. (d) An alternative representation of Riemann surface. (e) Refractive index profile and light ray trajectories. (f) A Gaussian beam impinges on the device at an incident angle of π/4 rad.

## 4. Specular reflector

To design a specular reflector, we utilize the same Non-Euclidean virtual space and the mirrors in Fig. 1(a). Moreover, we adopt two extra mirrors, depicted as two quarter-circles situated on the equators of both spheres, linking the red branch cut and the longitude circle mirror, as shown in Fig. 4(a). In this non-Euclidean virtual space, all light rays undergo specular reflection three times by the mirrors, preventing entry into the two half-spheres bounded by mirrors, which is illustrated by two typical blue and orange light rays in Fig. 4(a). Consequently, the two half-spheres also remain invisible. In the intermediate space of Fig. 4(b), following the same geodesic mapping from Fig. 1(a) to 1(b), the blue and orange light rays reflect twice off the mirrors and go back to the branch cut, appearing as if they have undergone specular reflection by the branch cut. Similarly, Fig. 4(c) illustrates the same Riemann surface via exponential mapping from Fig. 1(b) to 1(c), with inclusion of two segment of mirrors represented by two thick black solid lines connecting to the red branch cut, perpendicular to the three thick black solid parallel lines. An alternative representation of the Riemann surface in Fig.

4(d) demonstrates that the blue and orange light rays are spectrally reflected to the branch cut, with the additional reflection provided by two mirrors shaped like thick black quarter circles.

Utilizing the same dual-logarithmic mapping from Fig. 1(c) to 1(e), we obtain the specular reflector depicted in Fig. 4(e), characterized by an identical refractive profile to that of Fig. 1(e) and 3(e). The distinction lies in the inclusion of all mirrors depicted in Fig. 1(e) and 3(e). Parallel light rays are spectrally reflected towards the red circle-like line, which corresponds to the red branch cut in Fig. 4(c). The reflection effect for a Gaussian beam is illustrated in Fig. 4(f).

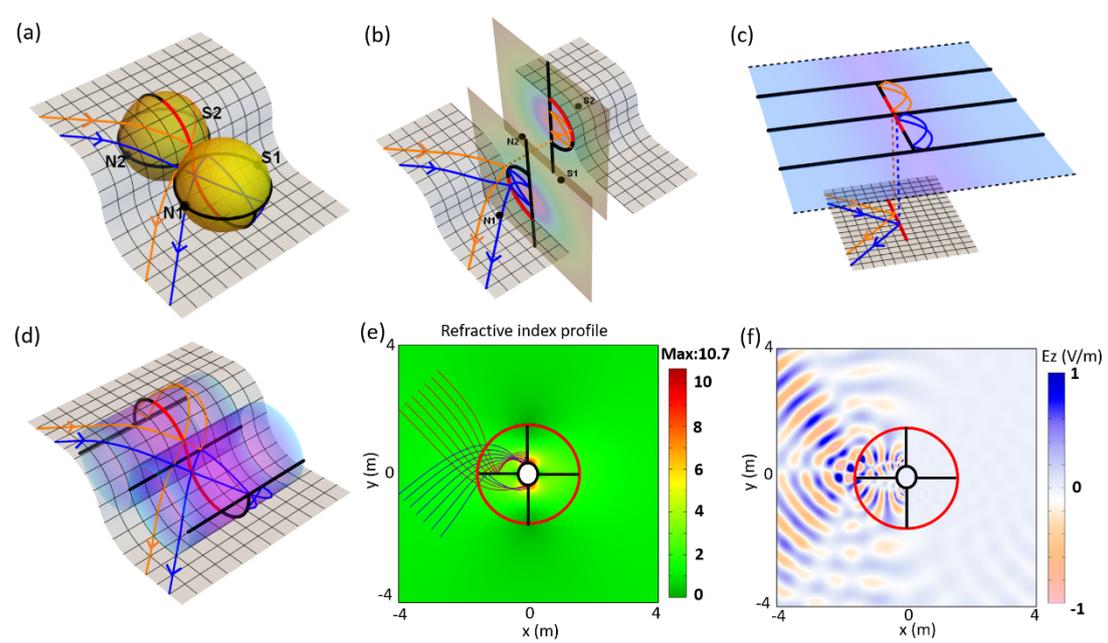

**Fig. 4. Specular reflector based on bi-spheres**. (a) Non-Euclidean virtual space. (b) Intermediate space after geodesic mapping. (c) Riemann surface. (d) An alternative intuitive representation of Riemann surface.(e) Refractive index profile and light ray trajectories. (f) A Gaussian beam impinges on the device at an incident angle of π/4 rad.

## 5. Conformal invisible device with bi-spheres in different sizes

Now, let's adjust the sizes of bi-spheres in the non-Euclidean space to design conformal invisible devices. As shown in Fig. 5(a), the radii of two spheres are $0.5r_0$ and $1.5r_0$ respectively, which differ from in Fig. 1(a). The two mirrors marked in black circles are positioned on two longitude circles of the two spheres. Consequently, the red branch cut is divided into two segments with a width ratio of 1:3. Here, two half-spheres with different radii remain invisible. In the intermediate space depicted in Fig. 5(b) following the same geodesic mapping from Fig. 1(a) to 1(b), the blue and orange light rays maintain their original direction upon exiting the branch cut. After the same exponential mapping from Fig. 3(b) to 3(c), its corresponding Riemann surfaces are illustrated in Fig. 5(c). There are three solid black lines representing mapped mirrors, with the middle one shifted to a quarter distance, different from that in Fig. 1(c). Therefore, two-sided mirrors are positioned at the distances of $0.5r_0$ and $1.5r_0$ from the two endpoints of the branch cut, respectively. The blue and orange light rays demonstrate the invisible cloaking effect in both Riemann surface depicted in Fig. 5(c) and its alternative intuitive representation in Fig. 5(d). After applying the same conformal dual-logarithmic mapping from Fig. 1(c) to 1(e), the refractive index profile of the invisible device ranges from 0 to 14.3, as shown in Fig. 5(e). The image of the middle mirror in Fig. 5(c) are mapped to two black curves touched the red circle-like curve in Fig. 5(e), and other two mirrors are mapped to form a closed black curve near the center. The propagations of parallel rays in Fig. 5(e) and of the Gaussian beam in Fig. 5(d) illustrate the cloaking effect, similar to those in Fig. 1(e) and 1(f), respectively. However, their eigenfrequencies differ due to the different sizes chosen according to Eq. (4). Consequently, cloaking effect is absent for this invisible device in wave optics regime.

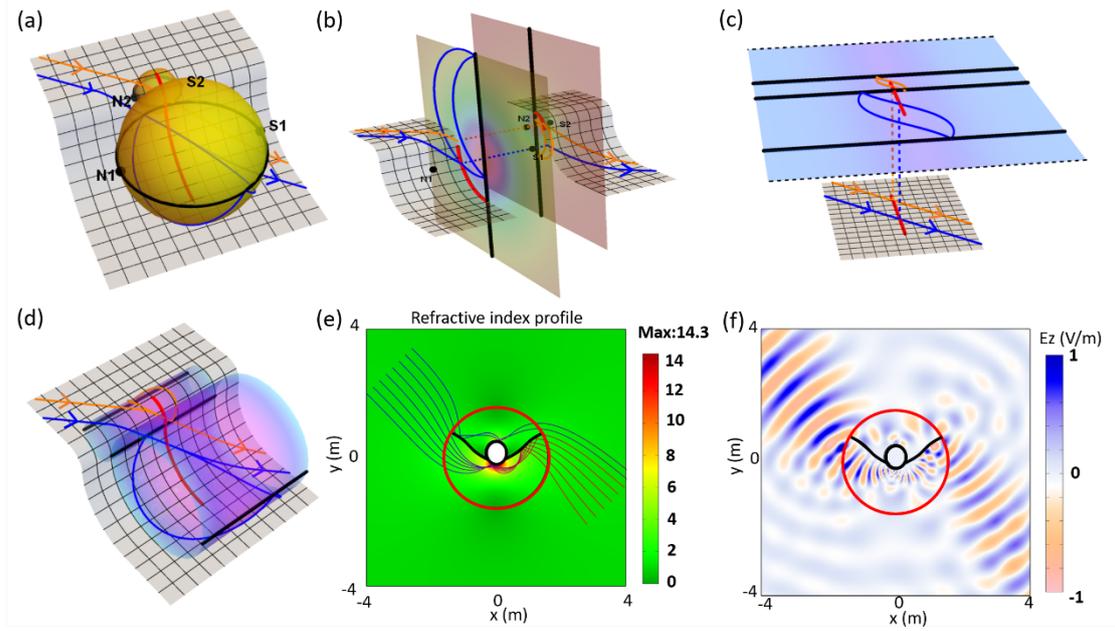

**Fig. 5. Conformal invisible cloak based on bi-spheres of different sizes**. (a) Non-Euclidean virtual space. (b) Intermediate space after geodesic mapping. (c) Riemann surface. (d) An alternative representation of Riemann surface. (e) Refractive index profile and light ray trajectories. (f) A Gaussian beam impinges the device at an incident angle of π/4 rad.

## 6. Conclusion and discussion

In conclusion, we further explore designing conformal devices with continuously varying refractive index profile in the framework of conformal transformation optics with geodesic mapping starting from non-Euclidean virtual space, therefore expanding the toolkit applications for practical applications. Through adjustments in the sizes of bi-spheres and the positions of mirrors, we demonstrate cloaking and reflection effects tailored for specific purposes.

Combined with the quasi-conformal method, our work holds promise for creating finite-size optical devices by incorporating slightly anisotropic media [9]. Even further, conformal transformation optics will greatly facilitate optical simulations to mimic challenging-to-observe cosmological phenomena, including black holes [42], gravitational lensing [43], and Einstein's Rings [44]. Our design strategy utilizing non-Euclidean spaces and Riemann surfaces may find broader application on their optical analogue with continuously varying refractive index profiles.

## ACKNOWLEDGMENTS

Our paper was supported by National Natural Science Foundation of China (grant No. 92050102, 12104012, 11904006 and 11804087), Science and Technology Program of Hubei Province (2022CFB553, 2022CFA012) and Educational Commission of Hubei Province of China (Q20211008).